\documentclass[fleqn,usenatbib]{mnras}

\usepackage{newtxtext,newtxmath}

\usepackage[T1]{fontenc}
\usepackage{ae,aecompl}
\usepackage{appendix}

\usepackage{float}
\restylefloat{figure}

\usepackage{graphicx}	
\usepackage{amsmath}	
\usepackage{subfig}
\usepackage{color}
\usepackage[utf8]{inputenc}
\usepackage[french,german,spanish,english]{babel}
\usepackage[normalem]{ulem}

\newcommand{\msol}{\rm M_{\odot}}

\newcommand{\fesc}{\rm f_{\rm esc}}

\newcommand{\hmpc}{$\rm{h^{-1}Mpc}$ }

\newcommand{\Msun}{\ensuremath{\mathrm{M}_{\odot}} }

\newcommand{\hmkpc}{$\rm{h^{-1}kpc}$ }

\title[Lyman-alpha opacities and radiative suppression]{Lyman-alpha opacities at z=4-6 require low mass, radiatively-suppressed galaxies to drive cosmic reionization.}

\author[P. Ocvirk]{
Pierre Ocvirk$^{1}$,
Joseph S. W. Lewis$^{1}$,
Nicolas Gillet$^{1}$,
Jonathan Chardin$^{1}$,
\newauthor
Dominique Aubert$^{1}$,
Nicolas Deparis$^{1}$ \&
\'Emilie Th\'elie$^{1}$
\\
$^{1}$Observatoire Astronomique de Strasbourg, Université de Strasbourg, CNRS UMR 7550, 11 rue de l’Université, 67000 Strasbourg, France\\
}
\date{Accepted 31/08/2021 for publication in MNRAS. Received YYY; in original form ZZZ}

\pubyear{2021}

\begin{document}
\label{firstpage}
\pagerange{\pageref{firstpage}--\pageref{lastpage}}
\maketitle

\begin{abstract}
  The high redshift Lyman-$\alpha$ forest, in particular the Gunn-Peterson trough, is the most unambiguous signature of the neutral to ionized transition of the intergalactic medium (IGM) taking place during the Epoch of Reionization (EoR). 
  Recent studies have shown that reproducing the observed Lyman-$\alpha$ opacity distributions after overlap required a non-monotonous evolution of cosmic emissivity: rising, peaking at z$\sim 6$, and then decreasing onwards to z=4. Such an evolution is puzzling considering galaxy buildup and the cosmic star formation rate are still continously on the rise at these epochs.
  Here, we use new RAMSES-CUDATON simulations to show that such a peaked evolution may occur naturally in a fully coupled radiation-hydrodynamical framework. 
  In our fiducial run, cosmic emissivity at z$>$6 is dominated by a low mass (${\rm M_{DM}}<2 \times 10^9 \msol$), high escape fraction halo population, driving reionization, up to overlap. Approaching z=6, this population is radiatively suppressed due to the rising ionizing UV background, and its emissivity drops. In the meantime, the high mass halo population builds up and its emissivity rises, but not fast enough to compensate the dimming of the low mass haloes, because of low escape fractions. The combined ionizing emissivity of these two populations therefore naturally results in a rise and fall of the cosmic emissivity, from z=12 to z=4, with a peak at z$\sim 6$. An alternative run, which features higher escape fractions for the high mass haloes and later suppression at low mass, leads to overshooting the ionizing rate, over-ionizing the IGM and therefore too low Lyman-$\alpha$ opacities.


\end{abstract}





\section{Introduction}
The epoch of reionization (hereafter EoR) starts when the first stars begin producing neutral hydrogen (HI) ionizing photons, resulting in growing ionized regions in the intergalactic medium (hereafter IGM). In its simplest form, the EoR begins with the formation of the first pristine, metal-free stars stars at redshifts as high as z=30, with star formation taking place to progressively larger metal-rich haloes. However, modelling this process is made complicated by the fact that describing the progress and sources of reionization involves a number of (poorly constrained) parameters, such as the minimum halo mass of star-forming galaxies and their star-formation efficiency.
Despite recent advances in observations and theory, the debate is still ongoing with respect to the nature of the sources powering reionization. Several studies have ruled out high-redshift active galactic nuclei as a main player, predicting them to contribute only a few percent to cosmic reionization \citep{mitra2018,matsuoka2018,kulkarni2019_agn}. Then, focusing on galaxies, several recent, full radiation-hydrodynamical (hereafter RHD) studies have predicted that galaxies should be able to reionize the Universe on their own, provided their stellar populations provide enough ionizing photons (\citep{rosdahl_sphinx_2018,ocvirk-codaii}. However, the jury is still out concerning what class of galaxies are the main drivers of cosmic reionization. Most studies favor low mass or intermediate mass systems \citep{wise_introductory_2019,katz_tracing_2019,lewis_galactic_2020}, contrasting with other studies favoring high mass galaxies as the most proficient reionizers \citep{naidu2020_oligarchs}. Even more uncertain are the properties of these galaxies: what is their ionizing escape fraction, how do they react to supernova feedback and irradiation from internal and external sources?

These recent theoretical studies, despite methodological differences, reproduce fairly well a number of high-redshift observables related to the EoR, in particular the UV galaxy luminosity function, the electron-scattering optical depth seen by the cosmic microwave background, and the timing of reionization, completing reionization between z=5.3-6 , such as in \cite{kulkarni_large_2019} (hereinafter K19). 
However, beyond the {\em timing} of reionization, one particular constraint that RHD EoR studies have left relatively untapped is the post-overlap average neutral fraction, i.e. the residual neutral fraction remaining once reionization is finished, as determined for instance by \cite{fan_constraining_2006} (hereinafter F06), and we offer below some elements of thought as to why this constraint has perhaps not received the attention it should in full RHD simulations.

First of all, it appears the post-overlap residual neutral fraction ${\rm x_{HI}^{res}}$ is surprisingly difficult to reproduce in simulations, 
and it is common to see deviations from \cite{fan_constraining_2006} values in the literature, even when the timing of reionization is well reproduced. This happens both using a post-processing approach  \citep{aubert_reionization_2010,bauer2015}, as well as in a full RHD framework, using our group's RAMSES-CUDATON code \citep{ocvirk_cosmic_2016,ocvirk_impact_2019,ocvirk-codaii}, but also other groups' codes \citep{petkova2011,zawada2014,so2014,aubert_emma:_2015,wu_simulating_2019}. In \cite{ocvirk-codaii}, for instance, the residual neutral fraction is too low because the ionizing rate is too high, and the Universe ends up too transparent. Some works seem to fare better than others in this respect \cite{rosdahl_sphinx_2018,katz_census_2018}, but the interpretation of the apparent better match obtained is complicated by the use of a reduced or variable speed of light formalism, which has been shown to strongly impact the residual neutral fraction ${\rm x_{HI}^{res}}$ \citep{gnedin_proper_2016,ocvirk_impact_2019,wu_simulating_2019,wu_imprints_2019}, as well as its timing \citep{bauer2015,deparis_impact_2019}. 

This now long-lasting difficulty in reproducing accurately the \cite{fan_constraining_2006} residual hydrogen neutral fractions in RHD simulations has led a number of authors to simply not show the evolution of the neutral hydrogen fraction in their papers any more, and instead show the evolution of the {\em ionized} fraction with redshift, or resort to using a linear scale, which easily hides any discrepancy on the residual {\em neutral} fraction.
Let us recognize, though, that this reserve in comparing simulation results with the neutral hydrogen fraction from e.g. F06 is partly well-founded: indeed, the neutral fraction ${\rm x_{HI}^{res}}$ is not a direct observable, but the result of the modeling of the opacities of the high-z Lyman-alpha forest. Therefore, the ${\rm x_{HI}^{res}}$ of F06 depends on assumptions regarding cosmology, the gas density distribution, and the radiation field the gas is exposed to.
It is therefore preferable to work directly with these opacities than with the residual neutral hydrogen fraction when comparing simulations and observations, and this requires computing pseudo-spectra of hydrogen transmission along lines-of-sight through the simulated volume.

{The latest and highest quality high-redshift quasar datasets seem to confirm a fairly late end to reionization, at  z$\sim 5.3$ \citep{bosman2021}, i.e. significantly later than the commonly assumed value of z=6 of F06. Moreover, K19 and \cite{keating_long_2019,nasir2020} have shown that the statistical properties of the high-z Lyman-alpha forest could be well reproduced if reionization finished by redshift z$\sim 5.3$}, but it also required a fine-tuned, non-monotonous comoving ionizing emissivity, with a marked {\em decrease} after z=6, in stark contrast with, e.g., the evolution of comoving star formation density, which is always  monotonously increasing at the relevant epochs, because dark matter haloes keep building up, as shown in \cite{ocvirk_cosmic_2016,ocvirk-codaii} and \cite{bouwens_reionization_2015}. This fine-tuning is found in different forms in a number of other works, e.g. \cite{chardin_calibrating_2015,chardin_large-scale_2017,chardin2018}. While allowing for exquisite fidelity in reproducing high-z spectra of the Lyman-alpha forest, this framework leaves the question of how this fine-tuning of the emissivity arises an open matter.
In full RHD simulations, such a fine-tuning of the co-moving ionizing emissivity is not possible, because it is intrinsically, self-consistently tied to the star formation of galactic haloes and their ionizing escape fractions, which are both an outcome of the simulation, and can not be tuned or modulated as the simulation runs. Without this fine-tuning, it is very difficult to reproduce correctly the post-overlap Lyman-alpha forest at z=4-6, even just in average opacity.


In order to address this, some works take the route of using an evolving escape fraction \cite{dayal_reionization_2020}, which results in a peak or plateau of stellar emissivity, and then a decrease. However, it may be desirable, as we do in this work, to go beyond a simple global escape fraction framework, because most numerical works point at the escape fraction being first and foremost a strong function of halo mass. Other studies, for instance, achieve good post-overlap neutral fraction by invoking dust formation, conveniently turning down galaxy escape fraction between z=8 and 6 \cite{gnedin2014,puchwein_consistent_2019}. Such a dust-related decrease in galaxy escape fraction may indeed happen, as shown in some simulations \citep{yoo_origin_2020}. However, is dust the only possible process allowing for a drop in cosmic  emissivity? Also, is there actually enough dust at high redshift and in the right places to account for this drop? We leave these considerations for another time, as it is sure to be a hotly debated topic for a while. 

We must also note that the need for a drop in cosmic ionizing emissivity may actually be the manifestation of missing / inaccurate physics in some other fields of the models: for instance, it is very hard for the studies quoted above to feature both the volumes relevant for Lyman-$\alpha$ forest modelling {\em and} the spatial resolution required to resolve Lyman-limit systems, which may play an important role in controlling post-reionization photon mean free path \citep{wu2021_accuracy_M1}. Indeed, a more abundant/potent population of photon sinks after overlap, even short-lived, may reduce the need for a strong drop in the cosmic ionizing emissivity \cite{cain2021}, or make a shallower drop acceptable. The impact, implementation, and in particular the evolution of this population of sinks is another hot debate and is outside the scope of this paper.

Yet another caveat possibly complicating our understanding of the post-overlap neutral fraction in simulations resides in potential artefacts of the M1 method used for radiative transfer \cite{aubert_radiative_2008}. Indeed, \cite{wu2021_accuracy_M1} showed that it could lead to some degree of over-ionization in absorbers, hence artificially reducing their ability to self-shield and/or play their role as photon sinks. A more accurate radiative transfer method could be expected, in principle, to yield a slightly more neutral IGM and therefore also reduce to some extent the need for a drop in cosmic ionizing emissivity. The quantitative details of this artefact, and in particular its impact on fully-coupled RHD or post-processing simulations of the EoR is yet to be determined.

Having carefully laid out these cautionary aspects, we now focus for the present study on the following premise: it stands that a rising, then decreasing ionizing emissivity around overlap, as in K19, appears to be required to obtain reasonable opacities in the high-z, post-overlap Lyman-$\alpha$ forest. Without dust, this is puzzling because the build-up of the galaxy population is ongoing at these epochs, and the cosmic star formation rate keeps on rising monotonically. 

If we take this requirement at face value, what could be the physical process(es) promoting the decrease of the ionizing emissivity? Can it be reproduced self-consistently in a fully coupled RHD simulation? These are the questions we set up to address in this paper. We will present a set of fully-coupled RHD simulations, displaying a strong radiative feedback which reduces star formation in low mass haloes. In the best-matching simulation, these low mass haloes are also the main drivers of cosmic reionization. Therefore, as their star formation is suppressed by progressing reionization, the average ionizing emissivity goes down. This suggests that external radiative suppression of star formation in low mass haloes is a crucial mechanism in reproducing the transmission properties of the Lyman-alpha forest during the EoR and after overlap. The paper plan is as follows. Section 2 details the simulations performed, Sec. 3, their analysis and our results, before concluding in Sec. 4.

\section{Simulations}

\label{sec:methods}

\begin{table}
\begin{center}
\begin{tabular}{lr}
\hline
\multicolumn{2}{c}{Cosmology (Planck14)} \\
\hline
Dark energy density $\Omega_{\Lambda}$  & 0.693 \\
Matter density $\Omega_{\rm{m}}$  & 0.307 \\
Baryonic matter density $\Omega_{\rm{b}}$  & 0.048 \\
Hubble constant $h={\rm H}_0/(100 \, {\rm km/s})$  & 0.677 \\
Power spectrum & \\
$\,$ Normalization $\sigma_8$       &       0.8288 \\
$\,$ Index $n$  & 0.963 \\
\\
\hline
\multicolumn{2}{c}{Setup} \\
\hline
Grid size   &	$1024^{3}$ \\
Comoving box size  &   11.8 Mpc (8 \hmpc)  \\
Comoving force resolution dx & 11.53 kpc \\
Physical force resolution at z=6 & 1.65 kpc \\
DM particle number ${\rm N_{DM}}$   &	$1024^3$ \\
DM particle mass ${\rm M_{DM}}$	 & 5.09 x $10^4$ \Msun \\
Average cell gas mass   & 9.375 x $10^3$ \Msun \\
Initial redshift $z_{start}$  & 150 \\
End redshift $z_{end}$   & 4.0 \\
\\
\hline
\multicolumn{2}{c}{Star formation}\\
\hline
Density threshold $\delta_{\star}$  & $50 \, \langle \rho_{\rm gas} \rangle$ \\
Temperature threshold $T_{\star}$ &	\\
\hspace{1cm } Fiducial & $2 \times 10^4$K \\
\hspace{1cm } Permissive & none \\
Efficiency $\epsilon_{\star}$  & \\
\hspace{1cm } Fiducial & $0.03$  \\
\hspace{1cm } Permissive & $0.02$  \\
Stellar particle birth mass $M_{\star}$  & 11732 \Msun \\
\\
\hline
\multicolumn{2}{c}{Feedback}\\
\hline
Massive star lifetime $t_{\star}$  & 10 Myr \\
\hline
\multicolumn{2}{c}{Supernova}\\
\hline
Mass fraction $\eta_{SN}$  & 20\% \\
Energy $E_{SN}$  &  $10^{51}$ erg \\
Metal yield $y$ & 10\% \\
\hline
\multicolumn{2}{c}{Radiation} \\
\hline
Stellar population model & BPASSv2.2.1 \\
Stellar particle escape fraction $\fesc^{sub}$ \\
\hspace{1cm} Fiducial & 1 \\
\hspace{1cm} Permissive & 0.45 \\
Effective photon energy$^{}$ $\bar{\epsilon}$   & 21.74 eV \\
Average HI ionization cross-section$^{}$ $\sigma_{N}$ & 2.69 x $10^{-22}$m$^2$ \\
Effective HI ionization cross-section$^{}$ $\sigma_E$ & 2.17 x $10^{-22}$m$^2$ \\ 
Speed of light $c$  & 299 792 458 m/s \\
\end{tabular}
\end{center}
\caption{Simulation parameters summary.}
\label{t:sum}
\end{table}

We use the numerical simulation code RAMSES-CUDATON, as presented in \cite{ocvirk_cosmic_2016,ocvirk-codaii}. The code couples RAMSES \cite{teyssier_cosmological_2002} with the radiative transfer module ATON \cite{aubert_radiative_2008}, resulting in a fully coupled radiation-hydrodynamics code for galaxy formation in a cosmological context. The radiative transfer module is optimized for running on Graphics Processing Units (GPUs), as described in \cite{aubert_reionization_2010}, taking advantage of their massive parallel computing power, hence the CUDA keyword. RAMSES handles the cosmological context, gravity, hydrodynamics, star formation and supernova feedback, including chemical enrichment, while ATON/CUDATON computes ionizing photon propagation and interaction with the hydrogen gas (photo-ionization, photo-heating and cooling processes). The code has been deployed on a variety of supercomputers, and has allowed us to produce the largest ever simulations of the EoR, Cosmic Dawn I and II (CoDa I and II), presented in \cite{ocvirk_cosmic_2016,ocvirk-codaii}. As an evolution beyond CoDa II, which was well calibrated with respect to a number of observables of the EoR, the setup for our new simulations  is similar in many ways, but differs in some key aspects. First of all, we are using here  
a 2 times higher spatial resolution, and 8 times higher mass resolution, in order to better resolve the range of physics  at play in haloes and in the IGM. 
CoDa I and II simulations were extremely heavy because of their very large size. Here, instead,  the box size is set to 8 \hmpc, to keep the computational cost in check, and the initial conditions are generated with mpgrafic \citep{prunet_mpgrafic_2013}, assuming the Planck cosmology \citep{planck14} given in Tab. \ref{t:sum}. The possible impact of the smaller box size will be discussed when necessary.
We outline below the main physical aspects of the code, highlighting the other differences with CoDa II physics.


To compute the H-ionizing emissivity of stellar particles as a function of their age and metallicity, we use BPASSv2.2.1 stellar population models \citep{eldridge_binary_2017} with binary stars. This is at variance with CoDa I and II, where emissivity was a step function, whereby a stellar particle would emit for 10 Myr and then be completely dark. We choose the  available initial mass function closest to \cite{kroupa_variation_2001} with slopes of -1.3 from 0.1 to 0.5 $\msol$ and -2.35 from 0.5 to 100 $\msol$. The mass fraction in supernovae is then set to $\eta_{\rm SN}=0.2$ as in \cite{rosdahl_sphinx_2018}. 

Chemical enrichment is accounted for, and allows us to track the increasing metallicity of galaxies gas and stellar content. We use a standard metal yield of $y=0.1$, in agreement with the BPASS model used, and the starting cosmic metallicity is 0. On average, the ionizing emissivity of a young stellar population decreases with increasing metallicity with this model.

Our stellar particles are relatively massive, close to $10^4$ $\msol$, i.e. a star cluster of intermediate mass. Such a cluster does not form its stars instantaneously, but over the course of a few Myr \cite{hollyhead2015,wall2020}. To account for this, we model the stellar particle as a population of constant star formation rate over 5 Myr, a timescale compatible with star cluster models of \cite{He2019,He2020} and compute the corresponding time-metallicity-dependent H-ionizing emissivity using the adopted BPASS models. We also compute the effective photon energy, average and effective ionization cross-sections given in Tab. \ref{t:sum}, following \cite{rosdahl_ramses-rt:_2013} Eqs. B3-B5, adopting for this an average absolute metallicity Z=$10^{-3}$ and integrating overs stars up to 10 Myr of age, after which the ionizing emissivity becomes too small to impact the radiative parameters significantly. A change in average metallicity or maximum age of integration affects the resulting photon energy and cross-sections at less than a few percent level, such that it does not affect our results significantly. We use the full speed of light in the radiative transfer module, for the propagation of radiation and its interaction with the Hydrogen gas, so as to avoid possible detrimental artefacts due to the reduced speed of light framework, as reported in \cite{deparis_impact_2019}. Of crucial importance for this study, and the Lyman-$\alpha$ forest in general, is the post-overlap residual neutral fraction, which can be strongly impacted by the use of a reduced speed of light, as shown in \cite{ocvirk_impact_2019}.

Finally, we will vary the sub-grid star formation model of the 2 main simulations in this paper.
One uses a strict threshold temperature for star formation ${\rm T_{\star}=2 \times 10^4 K}$, i.e. stars are allowed to form only in cells with lower temperature than ${\rm T_{\star}}$, as in CoDa I \citep{ocvirk_cosmic_2016}. This is a very common, widely used criterion, both in grid-based and Smoothed-Particle-Hydrodynamics codes, e.g. \citep{stinson2006,agertz2013}.
The other simulation does not use this criterion, and hence it is closer to the CoDa II setup \citep{ocvirk-codaii}. In the latter, star formation is therefore allowed in cells above the threshold density for star formation $\delta_{\star}$, no matter the temperature. Star formation is therefore more permissive. These 2 simulations will be referred to as "Fiducial" and "Permissive" in the rest of the paper. There is an underlying physical assumption which subtends these two models. The ${\rm T_{\star}=2 \times 10^4 K}$ limit is the highest temperature the gas can reach via photo-heating. Therefore, anything hotter is shock-heated, usually by supernovae. Therefore, the Permissive model, by allowing star formation beyond this limit temperature, implicitly assumes that an unspecified fraction of the unresolved cell gas is still in the form of cold gas that could form stars. On the contrary, the fiducial model implies that above ${\rm T_{\star}}$, the gas is in a shocked state, unable to form stars.
For the fiducial simulation, because star formation is less permissive than in CoDaII, the star formation efficiency was increased to $\epsilon=0.03$, and the particle escape fraction was also increased to 1, while the Permissive run used respectively 0.02 and 0.45 for these parameters. Both simulations ran down to z=4 to allow us to study the post-overlap state of the irradiated IGM.

\begin{figure*}[htbp]

  \includegraphics[width=1.\columnwidth]{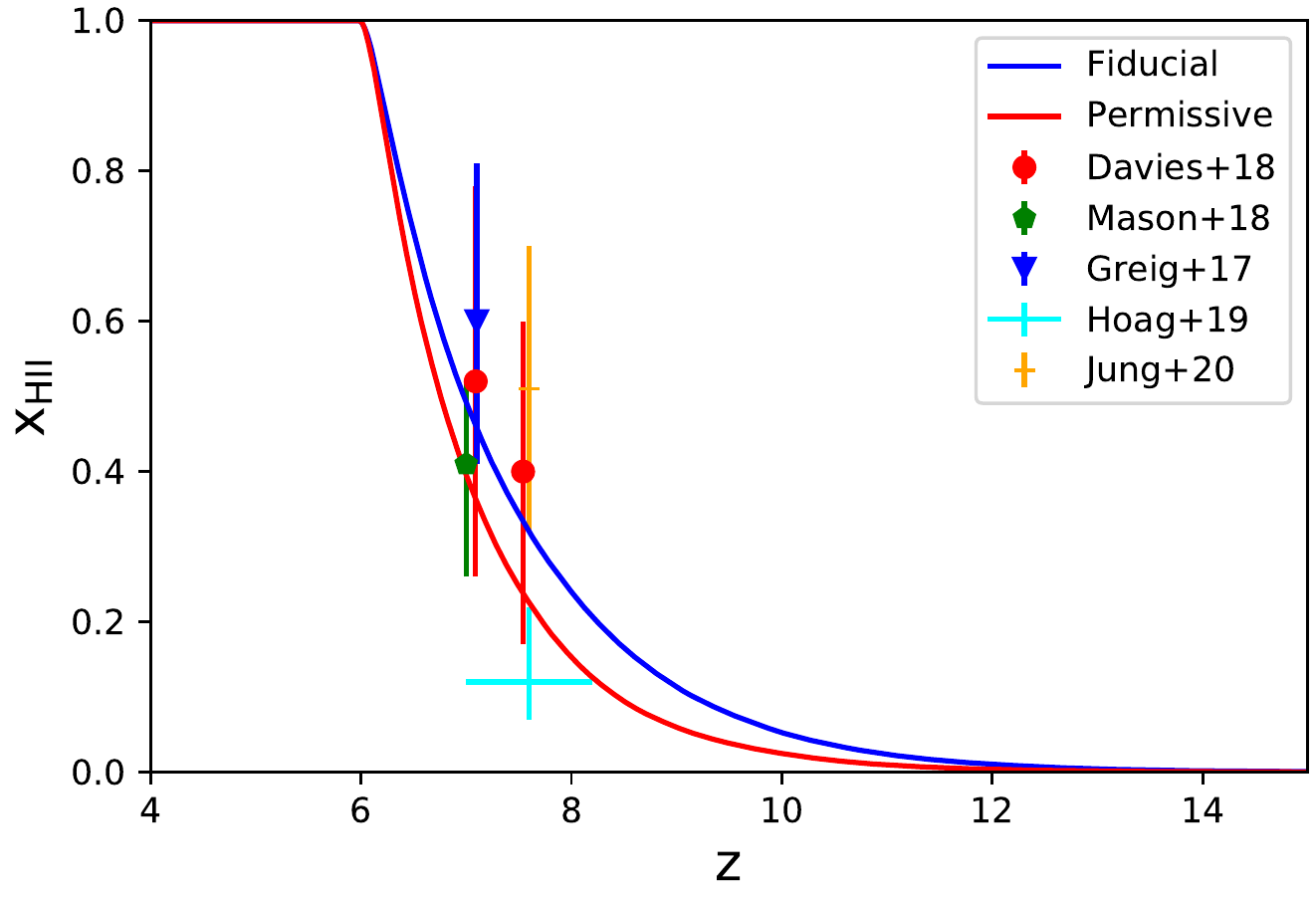}
  \includegraphics[width=1.\columnwidth]{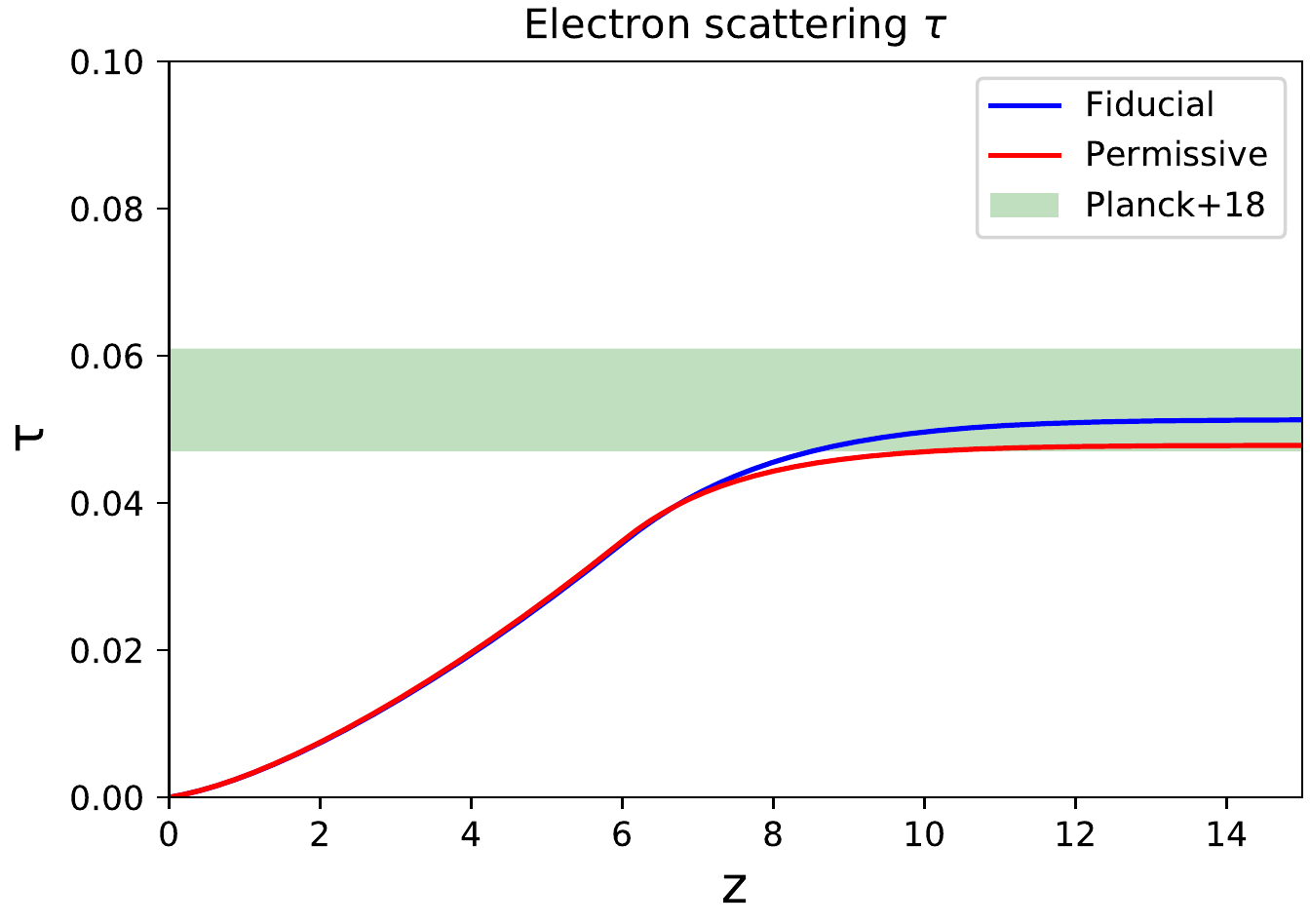}

\caption{Cosmic ionized fraction (left) and electron-scattering optical depth (right) from the simulations and comparison with observations.}
\label{fig:xhii_tau}
\end{figure*}

\section{Results}

\subsection{Ionized fraction and electron-scattering optical depth}

Both simulations are a good match to observational estimates of the cosmic ionized fraction \citep{greig2017,davies2018,mason_universe_2018,hoag2019,jung2020}, as shown in the left panel of Fig. \ref{fig:xhii_tau}.
The evolution of the ionized fraction in both runs is consistent with most of the observed data points. Although the fiducial run starts reionizing earlier because of the higher star formation efficiency and particle escape fraction $\rm{f_{esc}^{sub}}$, it then proceeds slightly slower than the Permissive run, which eventually catches up so that both runs are fully reionized at the same redshift z$\sim 6$. More precisely, if we define the reionization redshift z$_{\rm rei}$ as the redshift at which the ionized fraction reaches ${\rm x_{HII}}=0.99$, the fiducial (Permissive) run reionizes at z$_{\rm rei}=6.039 (6.032)$, so that the runs can be considered to effectively reach complete reionization at the same time. The headstart of the Fiducial run also transpires in the electron scattering optical depth, which is slightly higher than for the Permissive run as shown in the right panel of Fig. \ref{fig:xhii_tau}, although both runs are in good agreement with the observed values of \cite{planck2018}.

\subsection{Lyman-$\alpha$ forest and Gunn-Peterson trough}
Although box sizes of $\sim$ 40 \hmpc and beyond are preferred to study the Lyman-$\alpha$ forest's detailed properties, e.g. PDF, power spectrum, dark gaps and transmission spikes statistics \citep{bolton2009}, the {\em average} effective opacity, while containing less information than the PDF of effective opacities $\tau_{\rm eff}$, is less affected by box size.
Indeed, \cite{chardin_calibrating_2015} shows that the average effective opacity of their 40 \hmpc box is very similar to that of their 10 \hmpc box, as shown in their Fig. 12(b). Based on their plot, it seems unlikely that reducing the box size to the 8 \hmpc setup we use here would suddenly yield a strongly discrepant average $\tau_{\rm  eff}$. Moreover, Fig. 10 of \cite{bosman2018_10-50Mpc} shows that although the detailed PDF can be affected by a change of averaging length (from 10 to 50 \hmpc), the {\em average} effective opacity, read as $P(\tau)=0.5$, remains relatively stable, except around z=6, where the dispersion is the largest. We are therefore confident that our small box size does not strongly bias the {\em average} effective opacities we compute. 
However, as a precaution, we will refrain from analyzing $\tau_{\rm  eff}$ PDFs obtained from our simulations, as those may be more strongly affected and therefore potentially more difficult to interpret.

\begin{figure*}
\includegraphics[width=2.1\columnwidth]{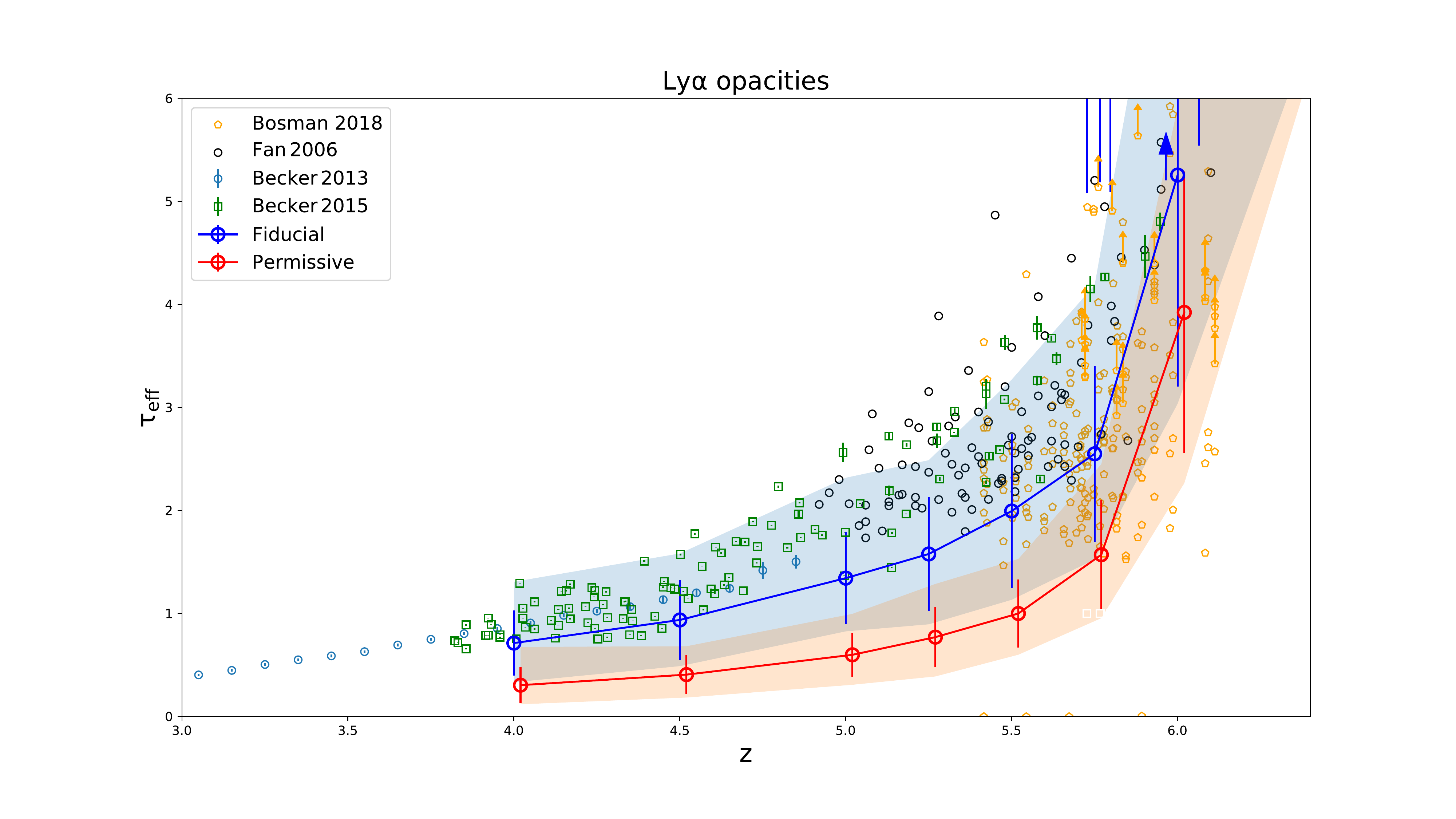}
\caption{Lyman-$\alpha$ effective optical depths predicted by our runs, and from observations. The solid blue (red) line shows the average of the effective optical depths of the Fiducial (Permissive) run, and their standard deviation as a vertical bar. The blue (orange) shaded area shows the extent of the 5-95 percentile of the distribution. The Permissive data points have been shifted by 0.02 in redshift, to improve lisibility.}
\label{fig:lya}
\end{figure*}

We consider 512 random lines of sight (hereafter LoS) of 8/h cMpc length through our volume. Along these LoSs, we compute the Lyman-$\alpha$ opacities of our simulated volume following \cite{chardin_calibrating_2015,chardin_large-scale_2017,chardin2018}, but without any adjustment or renormalization of the flux or neutral fraction values, just using the simulation's raw data.
We compute the average opacity of the k-th LoS as:
\begin{equation}
    \tau_{\rm eff}^{k}={\rm log \left( \frac{1}{N} \sum_i e^{-\tau_i^k}     \right)} \, , 
\end{equation}
where $\rm log$ is the Napierian or natural logarithm, $ \rm{  {\tau_i^k} }$ is the opacity of cell i on LoS k, and N is the number of cells spanning the 8 \hmpc LoS. The average opacity $\tau_{\rm eff}$ of the simulation's IGM is then just the linear average of the 512 $ \tau_{\rm eff}^{k} $.
The resulting average $\tau_{\rm eff}$ are shown in Fig. \ref{fig:lya} as a function of redshift, along with observations from \cite{fan_constraining_2006,becker2013,becker2015,bosman_new_2018}. The Permissive run's opacities are significantly lower than observed. The Fiducial run fares much better in this respect. 
The agreement is good just after overlap and down to z=5.5, after which the Fiducial run's average $\tau_{\rm eff}$ sits slightly below the observed cloud of points, before rejoining it around z=4.5. Despite the small offset at $\sim 5$, the Fiducial run clearly matches the observations better than the Permissive run, eventhough the reionization redshifts of both runs are equal. Remarkably, the Fiducial run is more opaque after overlap than the Permissive run, eventhough the Fiducial run starts reionizing earlier.
To understand this, we now turn to further quantities characterizing the gas opacity.

\subsection{Neutral fraction and ionizing rate}

\begin{figure*}

  \includegraphics[width=1.\columnwidth]{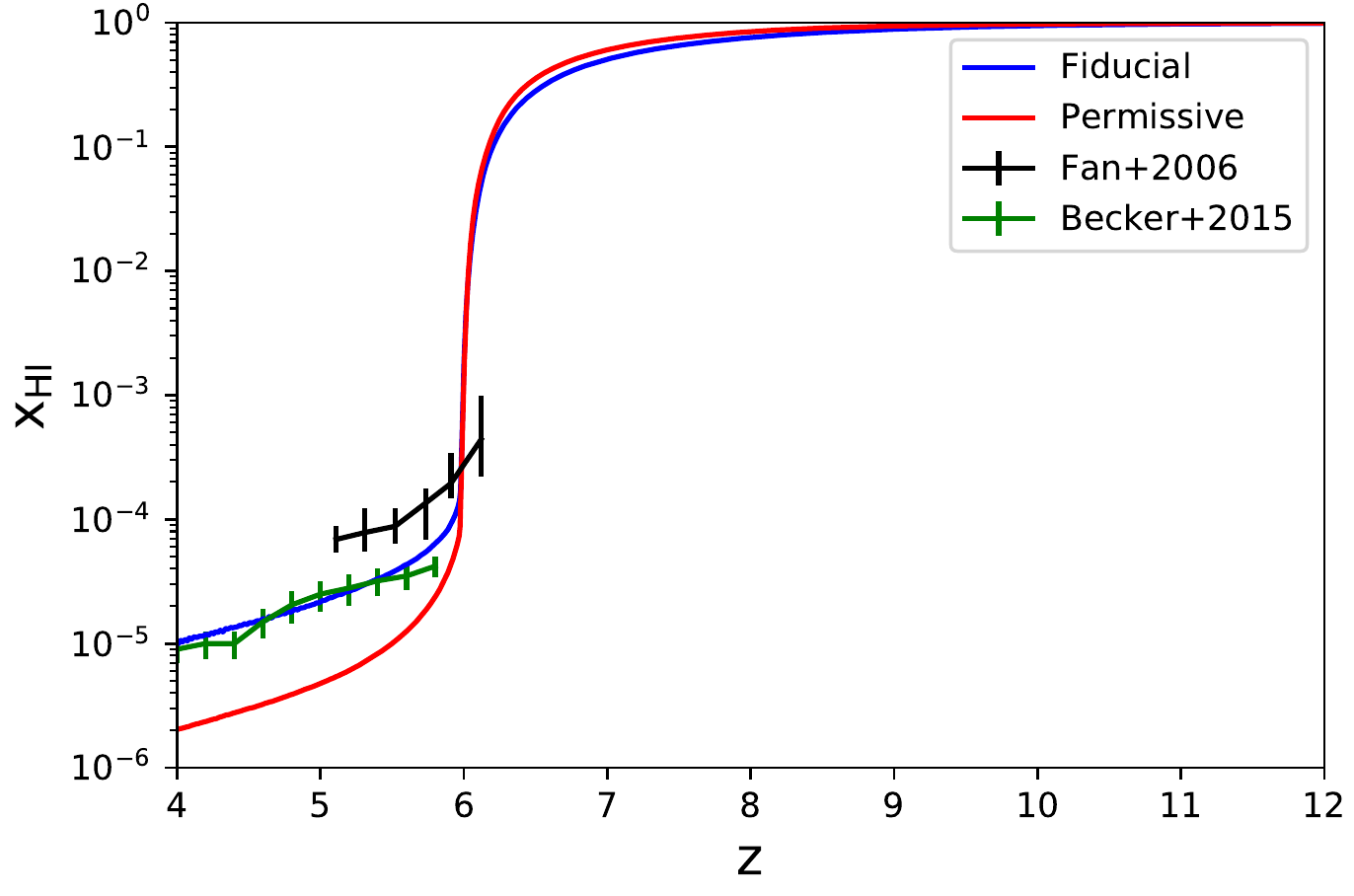}
  \includegraphics[width=1.\columnwidth]{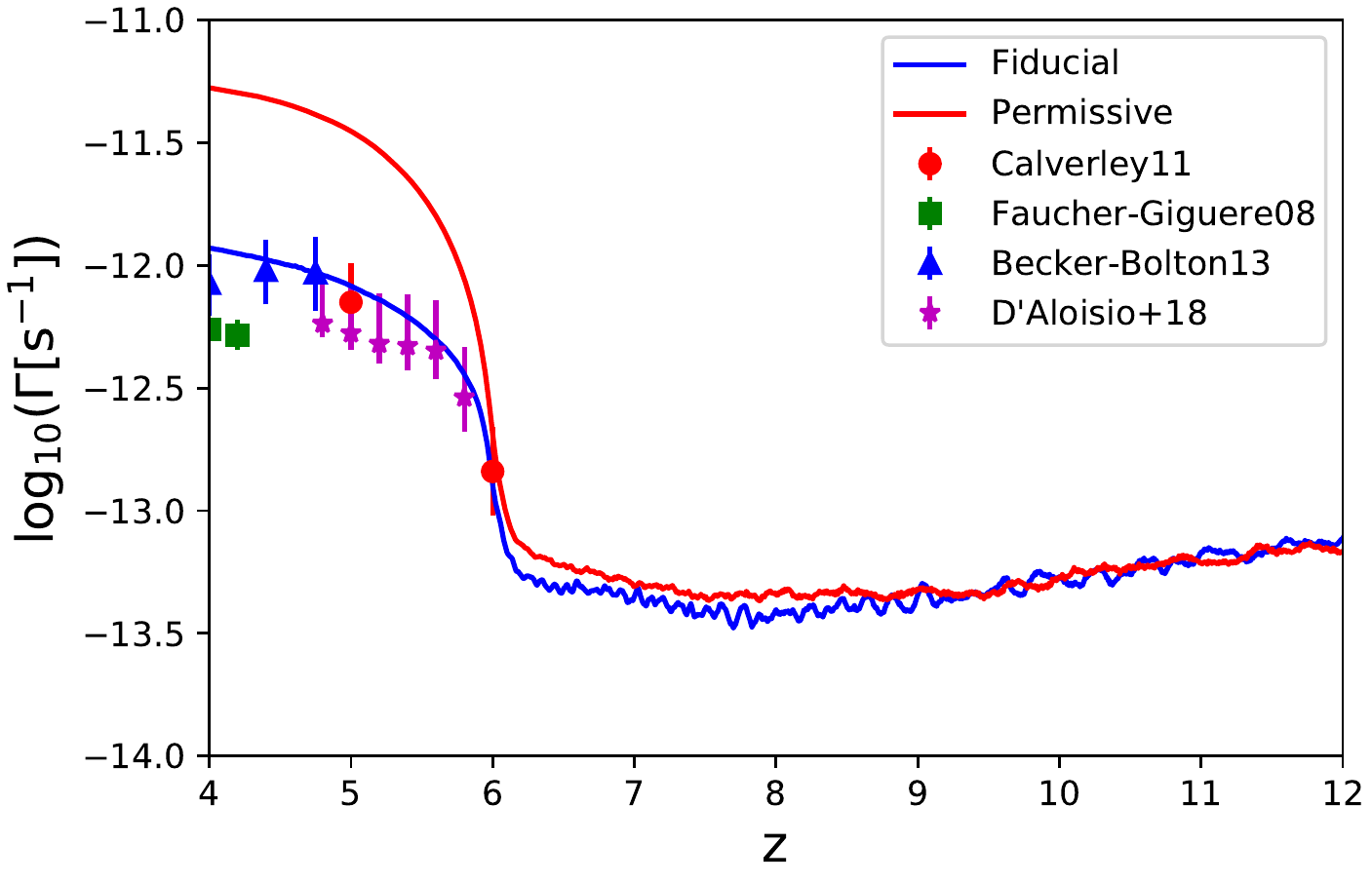}

\caption{Cosmic neutral Hydrogen fraction (left) and Hydrogen ionizing rate in ionized regions (right) from the simulations and comparison with observations.}
\label{fig:xhi_gam}
\end{figure*}

The Lyman=$\alpha$ opacities of the IGM are determined principally by the neutral Hydrogen density on the line of sight considered, which is, in turn, tightly connected to the ionizing rate $\Gamma$. In Fig. \ref{fig:xhi_gam} we show the hydrogen neutral fraction (left) and the hydrogen ionizing rate in ionized regions
(right) for both simulations, and existing observational constraints.
The \cite{fan_constraining_2006} neutral fractions are shown to guide the eye, but since they are derived from opacities, it is much better to compare directly the opacities, as we did in Fig. \ref{fig:lya}.
The fiducial run's post-overlap neutral fractions end up below the \cite{fan_constraining_2006} values agree better with the more recent data shown in Fig. 11 of  \cite{becker2015}.

Although reionization happens in both runs at precisely the same redshift, the neutral fractions of the Permissive run are smaller than the Fiducial run's by a factor $\sim 5$ at z=4, which explains their difference in opacity. 

The origin of this difference in post-overlap neutral fraction can be traced back to differences in ionizing rates, as seen in the right panel of Fig. \ref{fig:xhi_gam}. The evolution of the ionizing rate exhibits a sharp surge at overlap (close to z=6), and a slow saturation afterwards, as seen in most of the literature \cite{aubert_reionization_2010,ocvirk_cosmic_2016,rosdahl_sphinx_2018,ocvirk-codaii}. While, the Fiducial run is in good agreements with the observations of \cite{faucher-giguere2008,calverley_measurements_2011,becker-bolton2013,d'aloisio2018}, the Permissive run  overshoots them by a factor $\sim 5$, which explains the difference in neutral fractions.

In order to understand how this difference in ionizing rate arises, we turn to studying the collective escape emissivity of the star-forming haloes of our simulations.

\subsection{Galaxy populations and their ionizing photon output}

Here we use the formalism of \cite{lewis_galactic_2020}, who computed the photon budget of galaxies in the Cosmic Dawn II simulation \citep{ocvirk-codaii} to compare the behaviour of the Fiducial and Permissive runs in terms of halo ionizing emissivities. In all of this discussion of our results, we discuss only the dark matter halo mass of haloes and may therefore use the term mass without further distinction, since stellar mass will not be considered.

We use RAMSES's on-the-fly clump finder PHEW \citep{bleuler_phew_2015} to detect dark matter haloes with the parameters indicated in the paper. Then we use $R_{200}$ as a proxy to the virial radius, and compute each halo's ionizing escape fraction at $R_{200}$ via ray-tracing as in \cite{lewis_galactic_2020}. The halo ionizing escape emissivity is then the product of the intrinsical ionizing emissivity of its stellar population, taken from the BPASS model we used, and its halo escape fraction. We then compute the total escape emissivity ${\dot{{\rm n}}}_{\rm esc}$ of the full halo population as the sum of the escape emissivities of all the haloes in the simulation divided by the comoving volume of the box. For a number of reasons, ranging from imperfect halo detection at the low mass end to over-linking, some stars may lack an associated halo, 
as was already noted in \cite{ocvirk-codaii}. As a consequence, these  stars are not accounted for in individual halo emissivities. To account for them in our analysis, we correct the total escape emissivity $\rm{L_{tot}}$ using a correction factor defined as 
\begin{equation}
\rm{f_{cor}=SFR_{full box}/SFR_{haloes}} \, ,
\end{equation}
where $\rm{SFR_{full box}}$ is the SFR of the full computational volume, and $\rm{SFR_{haloes}}$ is the total SFR taking place in haloes. This correction factor varies from about 1.5 at z=10 to  1.1 at z=4.

\begin{figure}

  \includegraphics[width=0.99\columnwidth]{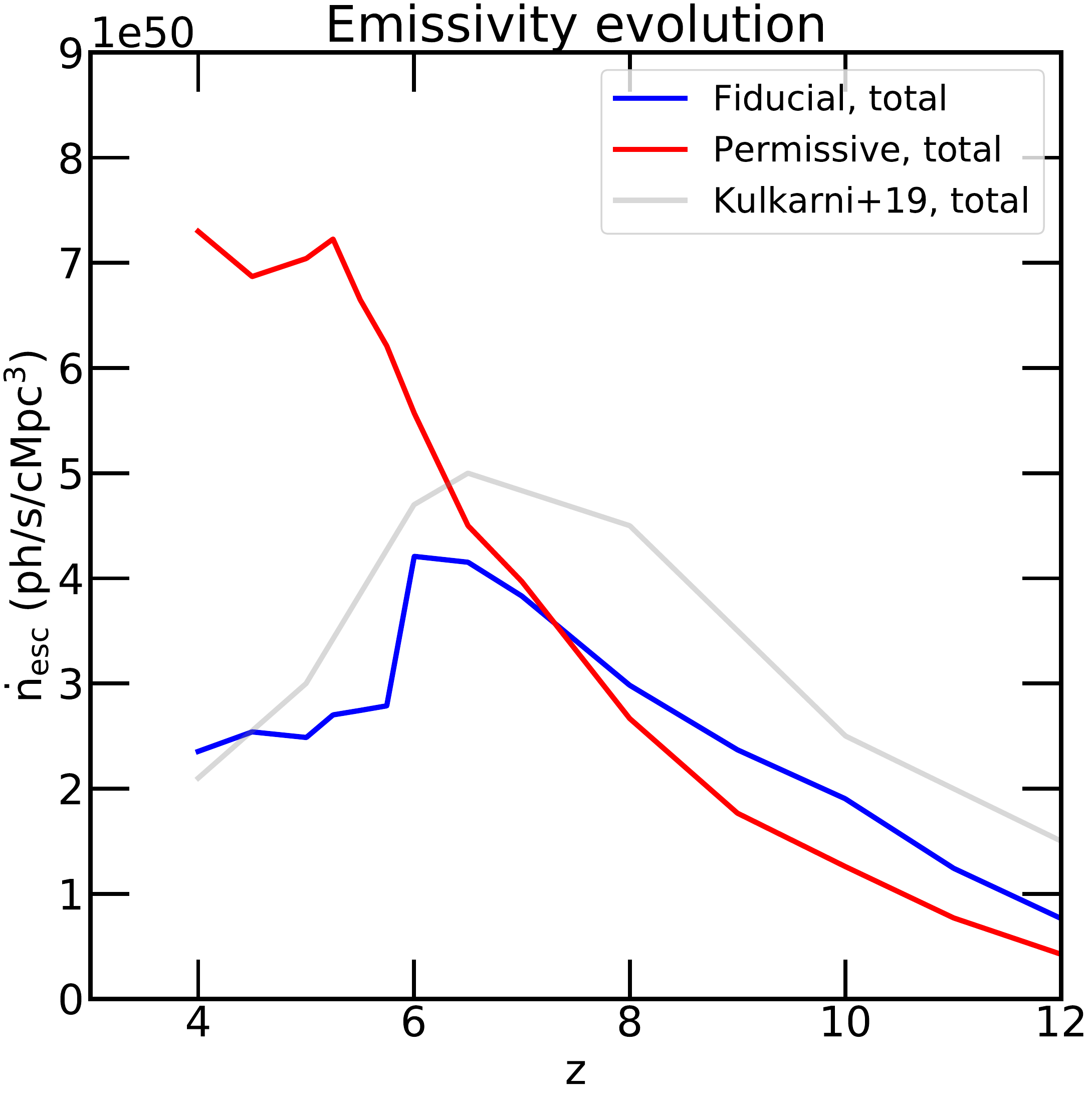}

\caption{Total Emissivities of our 2 simulations. The blue (red) solid line shows the total emissivity of the Fiducial (Permissive) run. The gray solid line shows the results of K19 for comparison.}
\label{fig:emtot}
\end{figure}

With this correction implemented, we show the total escape emissivity ${\dot{{\rm n}}}_{\rm esc}$, for our 2 simulations in Fig. \ref{fig:emtot}, as solid lines. The difference in evolution is striking: the Permissive run rises monotonously, and starts to level at z=5.5, whereas the Fiducial run's total emissivity increases up to z=6 and then {\em decreases} onwards. This difference in halo emissivity explains the different evolutions of the ionizing rates in the two simulations. We recall here that AGNs are not accounted for in our simulations, and that they may contribute significantly by the end of our runs, at z=4, as seen in \cite{dayal_reionization_2020}. We plan to explore this aspect once we have upgraded our code to model AGNs, in a future paper.

The overall evolution of the emissivity of the Fiducial run is reminiscent of that reported by K19, who found that a non-monotonic, rising then falling, emissivity, was required to reproduce the Lyman-$\alpha$ opacities at the end of the EoR. It may come as a surprise that, although the emissivity of our Fiducial run is lower than that of K19, it reionizes earlier. We caution however that our emissivities are perhaps not directly quantitatively comparable to theirs. Indeed, their spatial resolution is of $\sim 80$ \hmkpc comoving (about 8 times coarser than used here). At this resolution, massive haloes may still span a few grid cells in virial radius, and absorption of ionizing photons between the halo center and the virial radius, i.e. through the circum-galactic medium (hereafter CGM), mostly, happens, and is resolved by their radiative transfer prodedure. Therefore all of their emissivity does not necessarily reach the IGM. In contrast, thanks to the determination of our halos' escape fractions, our emissivity is a measure of what comes out of the halo population at the virial radius, and therefore directly into the IGM. Because of this difference, it is expected that K19 would require a higher emissivity than our runs to obtain a comparable reionization history for the same box size. On top of this, their box size is larger, and larger boxes take longer to reionize at fixed emissivity. Considering these 2 aspects, their lower reionization redshift despite higher overall  emissivity is not an issue. Therefore, we can not expect the agreement with K19 to be fully quantitative, but the qualitative agreement in the shape of the evolution of the emissivity, namely its rise and fall, is our main result here.

\begin{figure*}

\includegraphics[width=0.99\columnwidth]{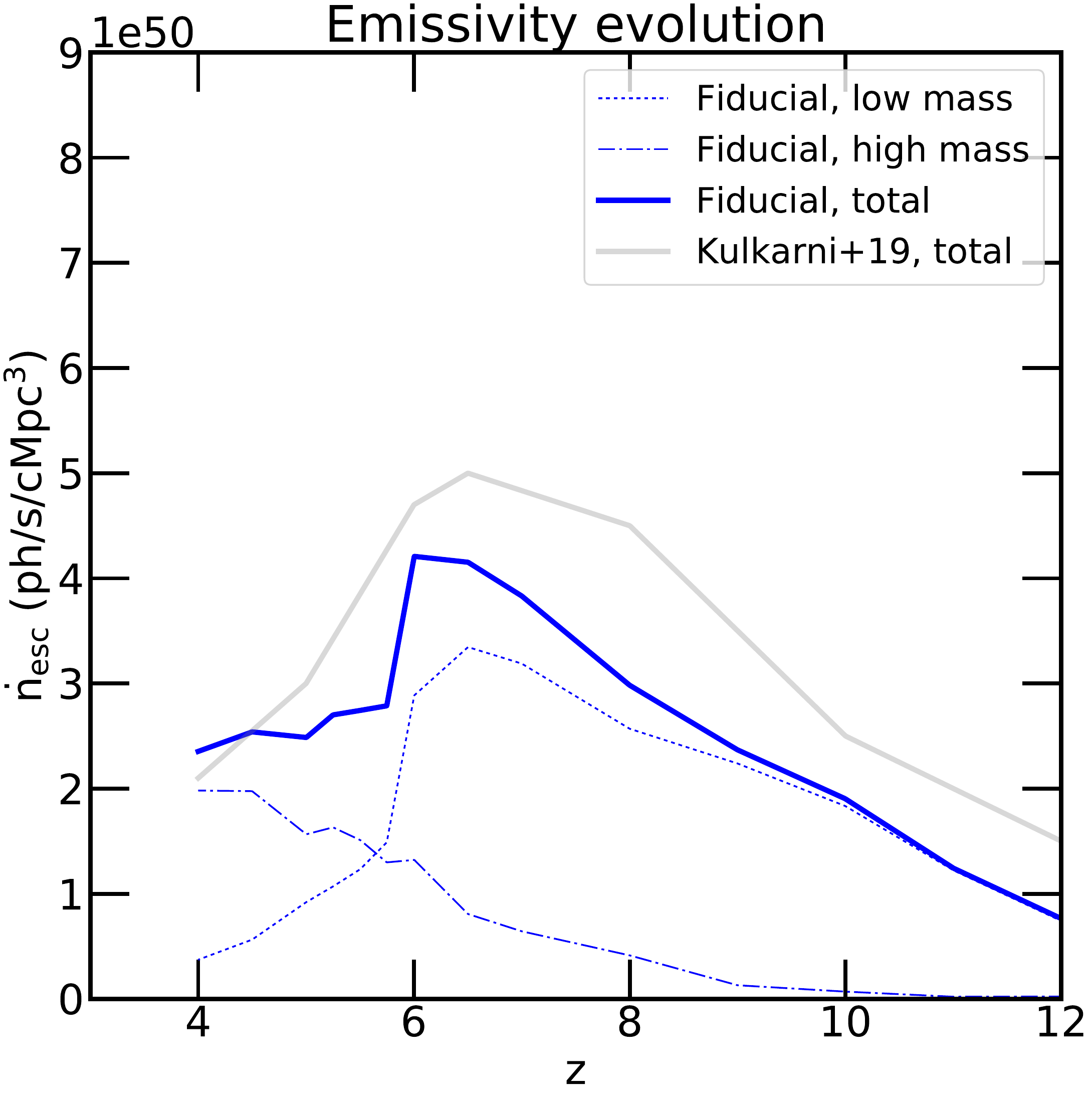}
\includegraphics[width=0.99\columnwidth]{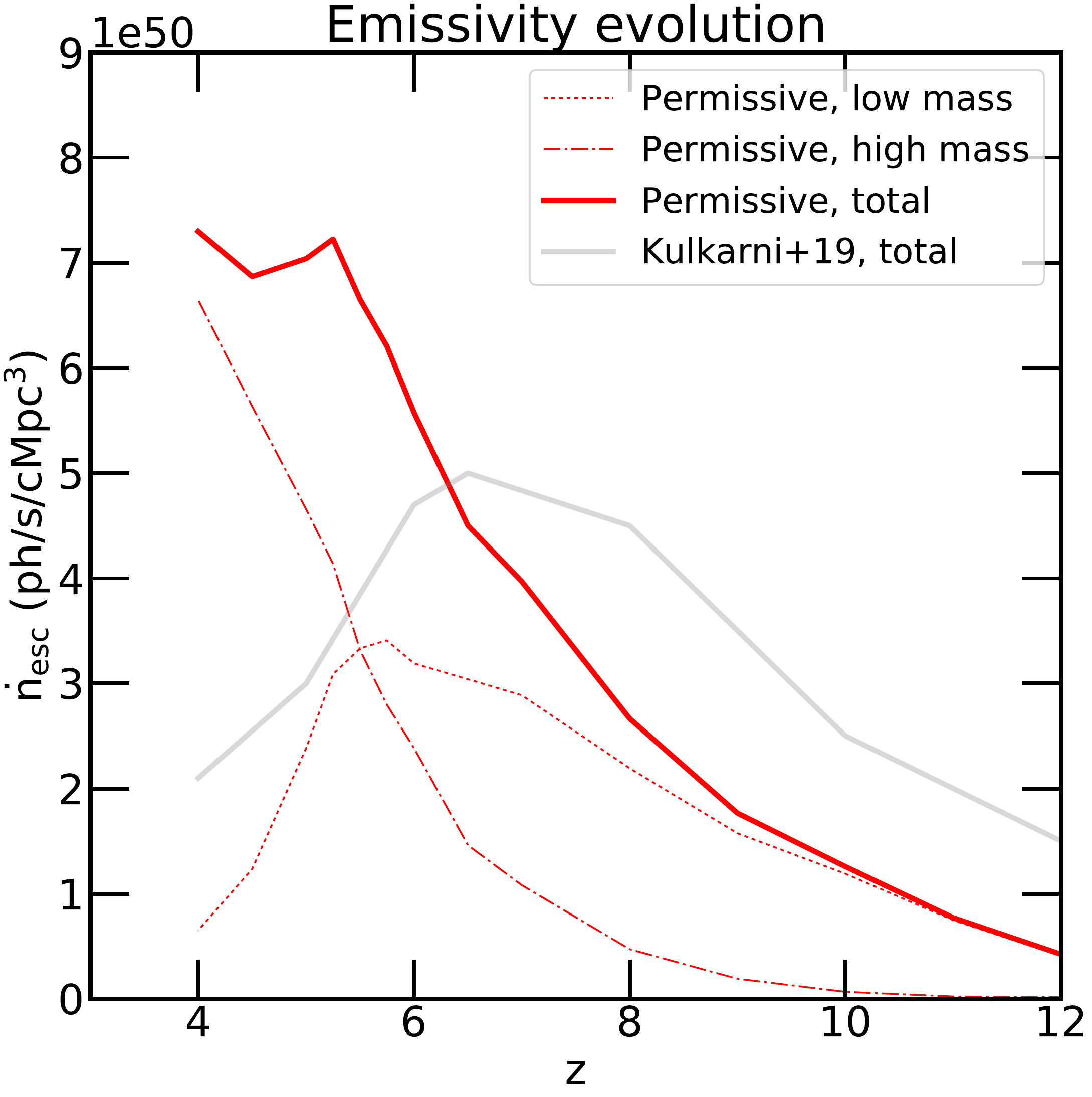}

\caption{Emissivities of our 2 simulations, detailed in two mass bins. The threshold between high and low mass haloes is set at $2 \times 10^9 \msol$. The blue (red) solid line shows the  emissivity of the Fiducial (Permissive) run. The gray solid line shows the results of K19 for comparison.}
\label{fig:emSP}
\end{figure*}

We now proceed to gain better insight into our results. In order to understand why the emissivities of the two runs are so different, we divided the total emissivity in a low mass and a high mass halo contribution, with a threshold at $2 \times 10^9\msol$. We use this threshold because it was determined to separate haloes sensitive to radiative feedback and haloes immune (in a star-foming sense) to the rise of the ionizing radiation field in our framework, as found in  \citep{ocvirk_cosmic_2016,dawoodbhoy_suppression_2018,ocvirk-codaii}, and in agreement with e.g. \cite{pawlik_spatially_2015,wu_simulating_2019}.
With this distinction, we see from Fig. \ref{fig:emSP} that indeed the contribution of low mass haloes decreases after reionization at z=6. The decrease occurs earlier for the Fiducial run than the Permissive run. 
Before z=6, though, the low mass haloes' contribution to the total emissivity is the main driver of cosmic reionization, as it is larger than that of the high mass haloes. The latter take over when the low mass population peters out. Eventhough they start to contribute later, because the massive halo population first needs to build up, once they are on the rise, their contribution increases monotonously. 
In the Fiducial run, the post-overlap decreasing total emissivity is the result of the demise of the low mass halo population, which the rise of the high mass counterpart does not compensate. 

While the details of the mechanisms involved in decreasing the emissivity in K19 are not explicitly given, our result sheds new light onto the origin of this peculiar evolution: it arises from the combination of a gradually suppressed population of low mass haloes driving reionization up to shortly before the overlap, when the contribution of high mass haloes starts to become significant. However, the latter contribution does not rise fast enough to compensate the quickly dimming low mass haloes, and the resulting total emissivity therefore decreases.

\begin{figure}

  \includegraphics[width=0.99\columnwidth]{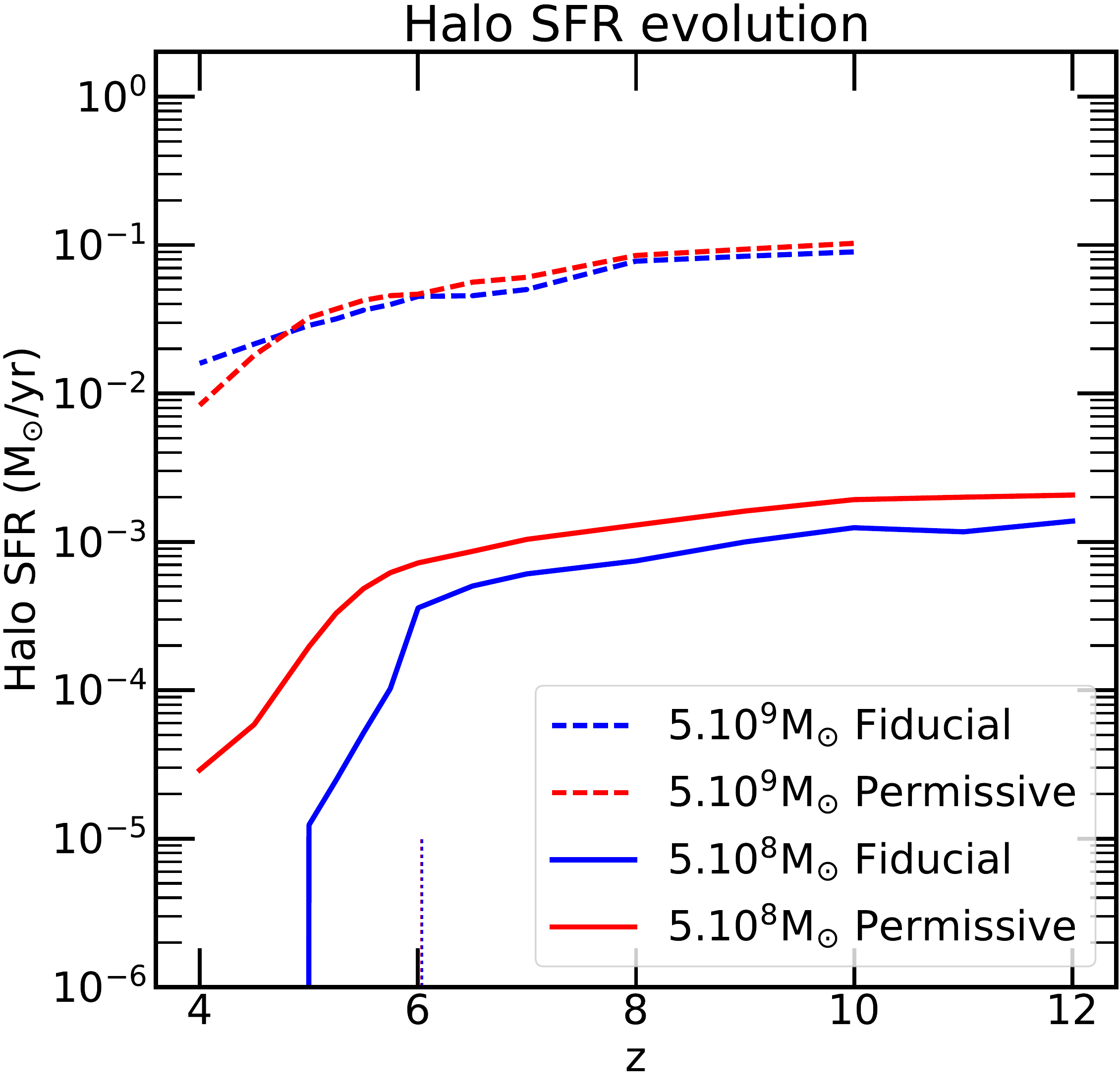}

\caption{Star formation rate of dark matter haloes in the range $2.5-7.5 \times 10^8 \msol$, i.e. centered on $5 \times 10^8 \msol$, for the Fiducial (blue, solid) and Permissive (red, solid) runs. The dashed lines show the SFRs of the $2.5-7.5 \times 10^9 \msol$ bin, i.e. centered on $5 \times 10^9 \msol$ bin. The short vertical dotted lines, blue and red, practically on top of each other, show the reionization redshift of both simulations.}
\label{fig:sfr}
\end{figure}

The later, and slightly shallower  decrease in emissivity of the low mass haloes seen in the Permissive run compared to the Fiducial run is expected, as it originates from a difference in SFR suppression. Such a difference was already reported in \cite{ocvirk-codaii}. It is further supported and confirmed by Fig. \ref{fig:sfr}, which shows the SFR of haloes in the low mass range ($2.5<{\rm M}/10^8 \msol < 7.5$) as a function of time (the masses quoted are instantaneous): the low mass haloes in the Fiducial run see their SFR suppressed more strongly than in the Permissive run. In contrast, $5 \times 10^9\msol$ haloes, belonging to the high mass group, see their SFR unaffected by the occurence of reionization. The shallow decrease they experience, in both runs, is of purely cosmological origin, and a similar behaviour was reported in cosmological simulations of galaxy formation without radiative transfer \citep{ocvirk_bimodal_2008}.
In summary, Fig. \ref{fig:sfr} confirms the suppression of SFR as the origin of the decrease of the low mass halo emissivity.

\begin{figure}

  \includegraphics[width=0.99\columnwidth]{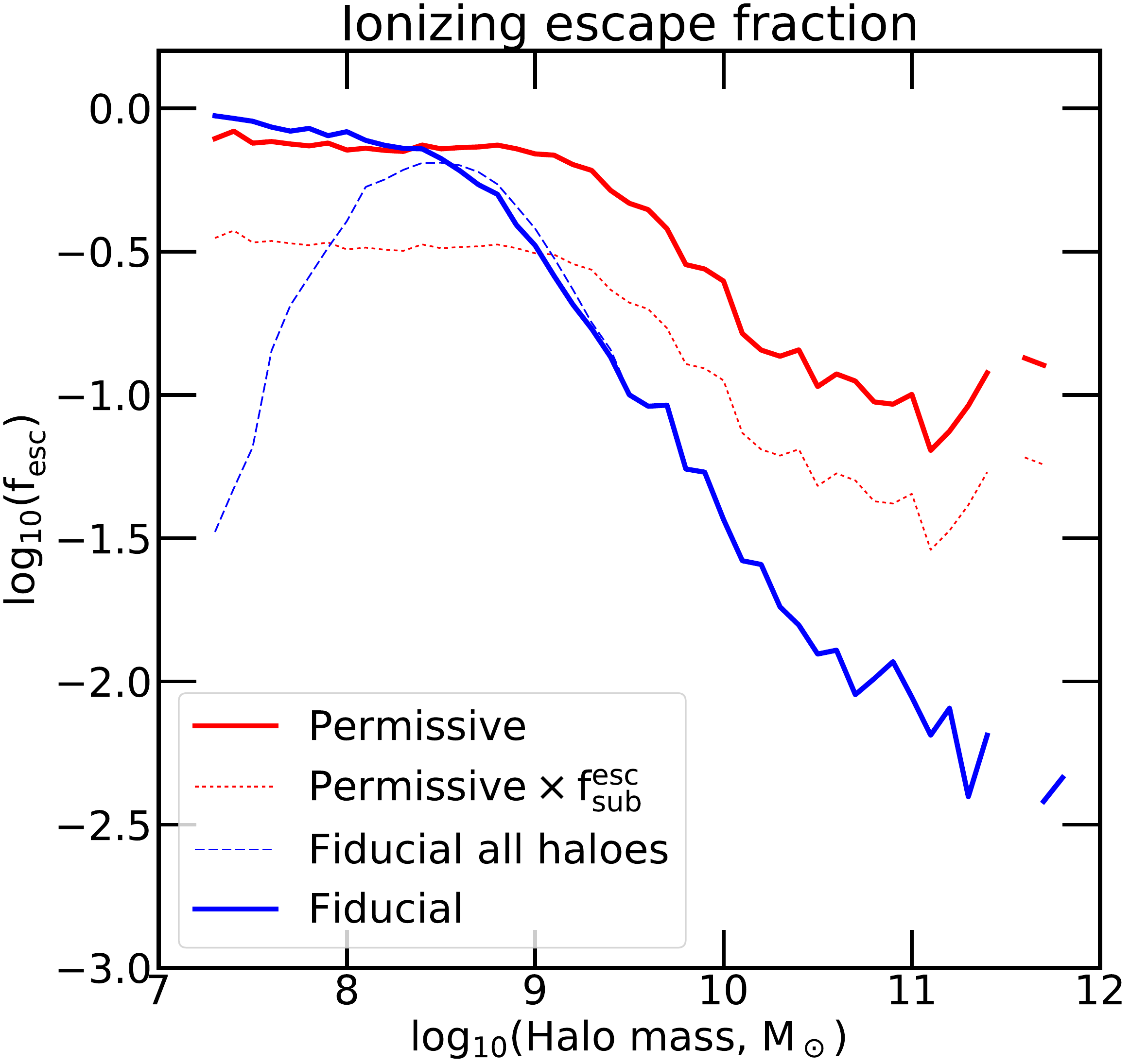}

\caption{Ray-tracing halo escape fractions measured in our 2 simulations. The solid lines show the averages of all star-forming haloes at all redshifts, for the mass bins considered. The blue dashed line shows the same but including {\em all} haloes, disregarding whether or not they are star-forming, for the Fiducial run. For the Permissive run, the dotted line shows the net halo escape fraction, i.e. the ray-tracing escape fraction multiplied by the stellar particle escape fraction, or sub-grid escape fraction, ${\rm f_{esc}^{sub}}=0.45$. For the Fiducial run, the net halo escape fraction is equal to the ray-tracing escape fraction, since ${\rm f_{esc}^{sub}}=1$.}
\label{fig:fesc}
\end{figure}

We note that the high mass haloes display significantly higher emissivities in the Permissive versus Fiducial runs. This is not expected from their SFR, which is fairly similar in both runs, as can be appreciated from the $5 \times 10^9\msol$ track of Fig. \ref{fig:sfr}. If the high mass haloes' SFRs are similar in the Fiducial and Permissive runs, then to allow the Permissive run's emissivity to be larger than the Fiducial run's, their escape fraction must be different. This is indeed the case, as shown by Fig. \ref{fig:fesc}, which represents the average ray-tracing escape fraction of all star-forming haloes, averaged over all redshifts, as a function of mass, for each run. Both saturate close to unity for low masses, but the escape fractions of the Fiducial run start dropping at lower masses than the Permissive run's and they drop more steeply. This difference is due to the Permissive run allowing stars to form in cells with temperatures potentially higher than $2 \times 10^4$ K, i.e. intrinsically more ionized and therefore transparent than the Fiducial run's.

Note, however, that in order for the Permissive run to complete reionization by z$\sim 6$, like the Fiducial run, we had to use a stellar particle escape fraction ${\rm f_{esc}^{sub}}=0.45$. Therefore, following the definitions in \cite{lewis_galactic_2020}, the net escape fraction for the Permissive run's is the ray-tracing escape fraction times ${\rm f_{esc}^{sub}}=0.45$, as shown by the dotted line in Fig. \ref{fig:fesc}.

The escape fractions obtained for the two runs are strikingly different, and one may wonder which of our runs is favored when comparing with the literature. In this respect, the most prominent feature of Fig. \ref{fig:fesc} is a decrease of escape fraction with mass, in agreement with most available numerical sutdies \citep{razoumov_ionizing_2010,yajima_escape_2011,wise_birth_2014,paardekooper_first_2015,katz_census_2018,kimm_escape_2014}, although the slope and extent of the decrease may vary, as can be expected given the range in resolution and modelling of these studies.

More specifically, the $\sim$unity escape fraction at low mass we obtain for both runs is compatible with the findings of \cite{razoumov_ionizing_2010,yajima_escape_2011,kimm_escape_2014}. At the high mass end, where the difference between the runs is largest, the escape fractions are in the $\sim 10$\% for the permissive run and in the few \% or sub-percent for the fiducial run. Therefore in this high mass regime, the permissive run is in better agreement with \cite{razoumov_ionizing_2010,kimm_escape_2014}, whereas the fiducial run is likely in better agreement with \cite{paardekooper_first_2015,wise_birth_2014,katz_census_2018}. We see that the escape fractions of our two runs span the range of predictions in the literature, meaning these previous studies are of relatively little guidance to discriminate which of our models is the more realistic.

Such comparisons are not straightforward: the methodology employed in the respective numerical simulations differs, but sometimes also the representation of the results can be misleading. For instance, we show in Fig. \ref{fig:fesc} our results for the Fiducial run (blue), in two flavours. The solid line shows the average escape fractions for {\em all star-forming haloes} (i.e. haloes must contain stars younger than 10 Myr), whereas the blue dotted line shows the average for {\em all haloes}, not considering whether or not they are star-forming. This results in a dramatic difference at the low mass end, which now features a sharp drop when considering all the haloes. This is remarkably similar to \cite{ma_fesc_2020}, who reported a comparable drop at low mass. This is due to the fact that star-forming haloes, harbouring active ionizing sources, are likely to be more ionized and therefore more transparent than non-star-forming haloes, where either no star was ever born or the gas has had time to recombine since the last star formation episode. Therefore the {\em all haloes} average escape fraction reflects to some extent the balance between star-forming and non star-forming haloes in our simulation. Since non star-forming haloes do not produce ionizing radiation, their escape fraction is irrelevant to the study of reionization, which is why we generally prefer to consider the star-forming haloes only statistics.
At the high mass end, the results of \cite{,ma_fesc_2020} are between 20\% and a few \%, i.e. they overlap with our two runs and the range between them. In the high mass regime, the star-forming versus all haloes distinction makes little difference because with increasing mass, all haloes become star-forming according to our definition.

{It was noted by \cite{becker_evidence_2018} that large scale metagalactic fluctuations in the ionizing UV background could exist around z=6. One may expect that such fluctuations favor a photon budget dominated by massive haloes. While our present simulations are too small to characterize such fluctuations, we note that this may corroborate the findings in our Fiducial run. Indeed, in the latter, at z=5 and below, the massive halo population dominates and contributes 2/3 of the total ionizing emissivity, with its contribution increasing further towards lower redshifts. Let us recall, however, that  UV background fluctuations need not be the only reason for the observed forest opacity fluctuations, and for instance, differences in reionization timing could be at play as well, as in K19. 
}

We also note, as a caveat, that the box size used here may not be sufficient to sample well the high mass end of the galaxy population. This may not be very important for the Fiducial run, because the high mass escape fractions are fairly low. However, for the Permissive run, escape fractions of high mass haloes are significantly larger, which makes them more important in the galactic photon budget, and therefore we may be missing their contribution or a fraction thereof because of the limited box size. For this reason, we refrain here from making strong statements about the photon budget, and in particular comparing directly our photon budget with \cite{lewis_galactic_2020} and literature, and we defer this comparison and deeper investigation to a later paper using Cosmic Dawn III, which will be 512 times larger in volume, when it will be fully available for analysis.




\section{Conclusions}

We have presented radiation-hydrodynamical simulations of the EoR with RAMSES-CUDATON, using two different sub-grid star formation models, tuned to reionize both at z=6, and analyzed their resulting Lyman-$\alpha$ opacities. The latter were then interpreted by looking at the emissivity of galaxy populations in both simulations. We find that the simulation best matching the observed quasar line-of-sight Lyman-alpha opacities is the Fiducial run, where star formation is allowed only below a gas temperature threshold of $\rm{T_{\star}}=2 \times 10^4$K, which promotes stronger radiative suppression, and escape fractions steeply declining with halo mass.
In this run, reionization is driven by a low mass halo population, up to overlap. At z=6, this population is radiatively suppressed due to the rising ionizing UV background, and its emissivity drops. In the meantime, the high mass halo population builds up and its emissivity rises, but not fast enough to compensate the dimming low mass one. The combined emissivity of these two populations therefore features a rise and fall, from z=12 to z=4, with a peak at z$\sim 6$. 
In contrast, our other run, with more permissive star formation, displays a more continued increase in emissivity, and as a result overshoots the observed ionizing rate and produces an overly transparent Universe after z=6, as shown by its low Lyman-$\alpha$ opacity. 
This suggests that a z=6 - peaked rather than continuously rising emissivity is required to avoid 
over-ionizing the Lyman-alpha forest after overlap. External radiative suppression of star formation in low mass haloes is a crucial mechanism in modulating the transmission properties of the Lyman-alpha forest during the EoR and after overlap.
The difference between the emissivities of the two runs occurs in two ways: the Fiducial run features stronger and earlier suppression at reionization, but also yields more steeply decreasing escape fractions with increasing halo mass, preventing the high mass halo population from compensating the low mass's dropping emissivity.This highlights the impact of the star formation sub-grid model not only on the SFR and its sensitivity to radiation, but also on the halo escape fraction. 

While the Fiducial run's Lyman-alpha optical depths fare better than the Permissive run's, there is still room for improvement. We foresee several directions to test and try for an improved agreement with observations. It is possible that aiming for a lower reionization redshift could help, as in K19 and  \cite{keating_long_2019}, however that may come at the cost of undershooting the Planck electron-scattering optical depth. In this direction, it is likely that using a larger volume would help us improve the significance of our study with respect to both observables. This aspect will be addressed with the upcoming Cosmic Dawn III simulation, which will be 512 times larger in volume, at the same spatial and mass resolution, and will therefore be a unique asset to extend current results.

We also plan to improve the physics of the simulation by modelling dust formation, and investigate the impact of this potentially important process on the evolution of the emissivity of the galaxies in our simulations. This will allow us to check which effect, radiative suppression of the high escape fraction population, or dust formation, as invoked in \cite{puchwein_consistent_2019}, contributes the most to the decrease of the cosmic emissivity. 


\section*{Data availability}
The data underlying this article will be shared on reasonable request to the corresponding author.

\section*{Acknowledgements}
This study was performed in the context of several French ANR (Agence Nationale de la Recherche) projects. PO acknowledges support from the French ANR funded project ORAGE (ANR-14-CE33-0016). The simulations in this study were performed on Jean Zay at Institut du Developpement et des Ressources en Informatique Scientifique (IDRIS) through several DARI and Grand Challenge allocations. The authors would like to acknowledge the High Performance Computing center of the University of Strasbourg for supporting this work by providing scientific support and access to computing resources. This work made use of v2.2.1 of the Binary Population and Spectral Synthesis (BPASS) models as last described in \cite{eldridge_binary_2017}. We thank the anonymous referee for the open discussion and impartial review provided, both of which have improved our paper.

\bibliographystyle{mnras}
\bibliography{Lya_Tsf.bib} 

\bsp	
\label{lastpage}
\end{document}